\documentclass[11pt,a4paper]{article}
\usepackage[T1]{fontenc}
\usepackage[utf8]{inputenc}
\usepackage{lmodern}
\usepackage{microtype}
\usepackage[a4paper,textwidth=160mm,textheight=240mm]{geometry}
\usepackage{amsmath,amssymb}
\usepackage{graphicx}
\usepackage{booktabs}
\usepackage{caption}
\usepackage{placeins}

\usepackage[numbers,sort&compress]{natbib}
\usepackage[hidelinks]{hyperref}
\usepackage{orcidlink}
\hypersetup{pdfstartview={XYZ null null 1}}
\usepackage{url}
\graphicspath{{figures/}}
\title{The optimal-transport cartogram:\\ world population as a Brenier map}
\author{Philipp Bogdan\,\orcidlink{0009-0009-7405-3555}\\ \small Department of Computing, Imperial College London\\ \small \texttt{pb825@ic.ac.uk}}
\date{September 2026}

\begin{document}
\maketitle

\begin{abstract}

A contiguous cartogram is a map whose area is proportional to a quantity such as population. The defining condition, that the Jacobian determinant of the deformation equals the density, is one equation for two unknown functions, so every cartogram method adds a tie-breaker, usually implicitly.
Optimal transport makes the tie-breaker explicit: among all density-equalising maps of the frame onto itself, take the one that moves the population least in the mean-square sense. By Brenier's theorem that map is the gradient of a convex potential, so it has no local rotation anywhere and cannot fold.
We compute this map for the world population of 2025 on a 4096 by 4096 Mercator grid (10 km cells) from the GHS-POP raster, using the fixed-point Monge–Ampère iteration of Benamou, Froese and Oberman with spectral Poisson solves, continuation in the population share and an ocean-only buffer, on a laptop GPU.
Against a Gastner–Newman diffusion cartogram of the same density we find the same transport cost to within 0.4 per cent and density errors of a few per cent for both, but a median local rotation of 8.8 degrees for diffusion against 0.01 for transport, and a median anisotropy of 6.30 against 3.93.
The construction extends to local refinement, by transporting a city's 100 m population onto the area measure the global map assigns to it, and to a semi-discrete counterpart: 8,192 Laguerre cells of 1.00 million people each.
Code, data provenance and every figure's inputs are public.

\end{abstract}

\section{Introduction}\label{introduction}

A cartogram redraws a map so that the area of each part is proportional
to a variable, usually population. Tobler traced the computer history of
the idea \citep{tobler2004}; Dougenik, Chrisman and Niemeyer's rubber-sheet
algorithm \citep{dougenik1985}, Gastner and Newman's diffusion method
\citep{gastner2004} and Gastner, Seguy and More's flow-based method
\citep{gastner2018} are the standard constructions, and Nusrat and Kobourov
survey the field \citep{nusrat2016}. For a continuous population density
\(\rho\) on a frame \(\Omega\), normalised to mean one, a contiguous
cartogram is a map \(T:\Omega\to\Omega\) with \(\det DT = \rho\): after
the deformation every unit of area holds the same number of people. This
is one scalar equation for the two components of \(T\). The diffusion
method resolves the ambiguity by following a physical process; the
flow-based method integrates a velocity field derived from one Poisson
solve; the rubber-sheet method iterates local expansions. None of these
states an objective that the final map minimises.

Optimal transport supplies one. Among all maps that push the population
measure \(\rho\,dx\) forward to the uniform measure on \(\Omega\),
choose the one minimising the transport cost
\[\int_\Omega \rho(x)\,|T(x)-x|^2\,dx .\] Brenier proved that the
minimiser exists, is unique \(\rho\)-almost everywhere and is the
gradient of a convex function, \(T=\nabla\varphi\) \citep{brenier1991};
Caffarelli proved that when both densities are bounded away from zero
and infinity on convex domains the map is smooth \citep{caffarelli1992}; Villani's
monograph is the reference \citep{villani2009}. Three properties follow for a
cartogram. The Jacobian \(DT = D^2\varphi\) is symmetric positive
definite, so its polar decomposition has no rotation: every small shape
is stretched along two perpendicular axes and never turned. The map is
monotone, so it cannot fold. And the total squared displacement of the
population is the smallest possible for any map with the same areas.

These properties are known in a neighbouring literature. Mesh adaptation
by equidistribution of a monitor function solves the same Monge--Ampère
equation, and Budd and Williams \citep{budd2009}, Budd, Russell and Walsh
\citep{budd2015}, Weller and colleagues on the sphere \citep{weller2016} and
McRae, Cotter and Budd \citep{mcrae2018} describe the resulting meshes as the
ones closest to the identity, gradients of a scalar potential and immune
to tangling. Zhao and colleagues used the same transport map to flatten
surfaces with prescribed area \citep{zhao2013}. In cartography we have not
found a published population cartogram computed as the Brenier map. The
flow-based method of Gastner, Seguy and More is described by its authors
as a flow whose equations of motion are easier to solve, not as a
transport map \citep{gastner2018}; Hennig's gridded population cartograms use
diffusion \citep{hennig2013}; Sargent's recent mesh optimisation minimises a
shape-distortion cost rather than displacement \citep{sargent2024}; Molchanov,
Rave and Linsen's integral-image method targets speed for time-varying
data \citep{molchanov2026}; Choi and Rycroft's density-equalising maps are
diffusion-based \citep{choi2018}. Of the 37 arXiv papers whose metadata
mention cartograms as of 12 September 2026, none uses optimal transport
\citep{repo}.

This note makes the construction and measures what it buys. Its
contributions are modest and stated plainly: the world population
cartogram as a Brenier map at 10 km resolution (Section 2 and Figures 1
and 2); a matched comparison with diffusion on the same density, which
shows that the two maps move the population equally far but differ in
rotation and shear (Section 3); a local refinement that transports a
city's 100 m population onto the area the global map gives it (Section
2.5 and Figure 5); and the semi-discrete counterpart, a power diagram of
humanity (Figure 6). The solver is not new; it is the fixed-point
iteration of Benamou, Froese and Oberman \citep{benamou2010} with fast Poisson
solves, a line that runs from Loeper and Rapetti \citep{loeper2005} through
Saumier, Agueh and Khouider \citep{saumier2015} to the back-and-forth method of
Jacobs and Léger \citep{jacobs2020}.

\section{Method}\label{method}

\subsection{The population measure}\label{the-population-measure}

The population raster is GHS-POP R2023A, epoch 2025, at 30 arc seconds,
which disaggregates census counts onto satellite-detected built-up area
\citep{ghspop2023}. Counts, never densities, are re-binned exactly onto a
\(W\times W\) Web-Mercator grid cut at latitude 85.05 degrees, so no
area correction is needed anywhere: each cell holds the people whose
coordinates fall in it. The frame's vertical edges are placed at
longitude 168 degrees west, in the Bering Strait, so that the Americas
lie at the western edge and the Pacific at the eastern edge and no
populated land is cut. The density is the count per cell after Gaussian
smoothing with \(\sigma\) = 30 km at the equator (3.1 cells at \(W\) =
4096), then normalised to mean one. Two constants complete the measure.
A floor adds a uniform density equal to a fraction \((1-s)/s\) of the
mean everywhere, where \(s\) is the share of the frame's area that the
population itself commands; \(s\) = 0.95 is the classic five per cent
floor. For the pictures we use a second, ocean-only version instead:
land keeps \(s\) = 0.999, and the ocean cells receive one uniform
density chosen so that the ocean occupies 20 per cent of the frame after
the map. The land metric does not depend on the ocean share; the ocean
is a margin that keeps the continents apart, a device every printed
cartogram atlas relies on \citep{hennig2013}.

\subsection{The transport map}\label{the-transport-map}

Write \(T(x) = x + \nabla\psi(x)\). The transport condition
\(\det DT = \rho\) is the Monge--Ampère equation
\(\det(I + D^2\psi) = \rho\), with the frame mapped onto itself. The
walls are realised by even reflection: \(\rho\) is extended
symmetrically to a torus of twice the side, the periodic problem is
solved with the fast Fourier transform, and by symmetry the map
preserves the square and its normal displacement vanishes on the walls.
In two dimensions
\((\Delta\varphi)^2 = |D^2\varphi|^2 + 2\det D^2\varphi\) for
\(\varphi = |x|^2/2 + \psi\), so the equation is equivalent to
\(\Delta\varphi = \sqrt{|D^2\varphi|^2 + 2\rho}\). Benamou, Froese and
Oberman's first method iterates this identity: evaluate the right-hand
side at the current iterate, solve one Poisson equation for the next
\citep{benamou2010}. We damp the iteration by one half and keep the best
iterate by residual. Two implementation details matter at this size.
First, the state is the Fourier transform of the right-hand side
\(f = \Delta\psi\), never the potential: the Hessian entries are
recovered as
\(D_{ij}\psi = \mathcal{F}^{-1}[(k_i k_j/|k|^2)\,\mathcal{F}f]\),
bounded multipliers applied to an order-one field, so single precision
on the GPU loses nothing, whereas differencing a potential of order
\(W^2\) pixels squared does. Second, the map is reached by continuation
in the share: the problem is solved at \(s\) = 0.95, 0.98, 0.99, 0.995
and 0.999 in turn, each warm-started from the last, after a
coarse-to-fine pass from 512 cells. With the spectral formulation the
continuation is insurance rather than a necessity: a direct solve at
\(s\) = 0.999 on a 2048-cell grid converges to a residual of 0.0106
against 0.0059 with continuation, with the same density error, in a
quarter of the time; every map reported here used the continuation. The
corner mesh of the output is \(x + \nabla\psi\) evaluated spectrally at
cell centres and averaged to corners. Cells whose signed area is not
positive are counted as folds; they are a discretisation failure of the
potential's convexity at the ocean creases, where the map compresses
cells far below one pixel, and they are reported, not hidden. For
drawing, the displacement is smoothed in folded cells that carry less
than a tenth of the mean population; populated folds would be left as
they are.

\subsection{The diffusion baseline}\label{the-diffusion-baseline}

Our comparison method is our own implementation of Gastner and Newman's
diffusion cartogram \citep{gastner2004} on the same grid, walls and density:
the heat equation is solved spectrally on the reflected domain, the mesh
corners are advected with velocity \(-\nabla\rho/\rho\) by fourth-order
Runge--Kutta with adaptive steps and a per-step displacement cap, from
\(t_0\) = 0.01 cells squared until the density is uniform to one part in
a thousand. One choice deserves a warning. Starting the flow at \(t_0\)
= 0.5 cells squared, which is natural when the raster is not
pre-smoothed, equalises the density diffused by that amount rather than
\(\rho\) itself and inflated the density error to about 9 per cent at
\(W\) = 1024 while leaving every other metric unchanged; with the
pre-smoothing used here the early start is safe and the error falls to
the level of the transport solve. The flow-based method of Gastner,
Seguy and More \citep{gastner2018} was not benchmarked: its reference
implementation is polygon-based, and our own gridded reimplementation
did not reproduce the published shape quality, so we do not report it.

\subsection{Measurements}\label{measurements}

All measurements are made on the raw corner mesh with the population per
source cell as weight. Density error is the population-weighted
distribution of \(\log(\rho/A)\) where \(A\) is the warped cell area; we
report its 5th and 95th percentiles as percentages. Anisotropy is the
ratio of the two singular values of the cell Jacobian. Rotation is the
angle of the orthogonal factor in the polar decomposition of the same
Jacobian, in degrees. Transport cost is the population-weighted mean of
\(|T(x)-x|^2\) in units of the frame width squared; it is the objective
that transport minimises. Shape error is the per-country Procrustes
residual, the root-mean-square distance between the warped outline of a
country and its best similarity-transformed original, normalised by the
original outline's root-mean-square radius, averaged over the countries
with at least one million people with population weights; the outlines
are Natural Earth 1:50m \citep{naturalearth}. Folds are cells of non-positive
area. Timings are wall-clock seconds for the solve on an Apple M4 Mac
mini with 16 GB of memory, PyTorch 2.12 on the Metal backend.

\subsection{Local refinement}\label{local-refinement}

The global map equalises the smoothed density at 10 km; inside a city it
inflates the city as a whole and leaves its blocks unequal. Let
\(\Omega_w\) be a two-degree window around a city and \(\mu\) its
population on the 3 arc-second raster, smoothed at 300 m. The global map
\(T\) assigns the window an area measure
\(\nu = (\det DT)^{-1}\circ T^{-1}\) pushed back to \(\Omega_w\), that
is, the area each source cell already receives. Solve the transport
problem on the window from \(\mu\) to \(\nu\): a map
\(L:\Omega_w\to\Omega_w\) with \(\det DL = \mu/(\nu\circ L)\). Then
\(T\circ L\) has Jacobian \(\mu\) on the window, so the city's blocks
get their own areas, and because \(L\) maps the window onto itself the
image \(T(\Omega_w)\) is unchanged and nothing overlaps a neighbour. The
same spectral solver takes the non-uniform target, sampling \(\nu\) at
the current map each iteration, with continuation from \(\nu^{1/2}\) to
\(\nu\). The composed map is no longer a Brenier map of \(\mu\)
globally, so the rotation-free property holds only up to the
composition; this is a rendering-scale device, not a claim of global
optimality. Carroll and Moore nested cartograms by user focus with the
diffusion method \citep{carroll2008}; nesting by data resolution follows the
same idea.

\subsection{The semi-discrete
counterpart}\label{the-semi-discrete-counterpart}

Replacing the uniform target by \(N\) point masses of equal weight gives
semi-discrete transport: the population measure is partitioned into
\(N\) Laguerre cells of exactly equal population, each a convex polygon,
the plane itself undeformed \citep{merigot2011}. We use pysdot \citep{pysdot}
with \(N\) = 8,192 on a 2048-cell grid, Newton's method on the weights
with continuation in the smoothing of the density from 32 cells to none,
because the Newton step needs every cell non-empty at the start.
Cohen-Addad, Klein and Young used balanced power diagrams of population
for redistricting \citep{cohenaddad2018}; here the same object is a picture of the
world.

\section{Results}\label{results}

Figure 1 is the construction in one pair of pictures: one dot per
million people placed on the Mercator frame by systematic sampling of
the raster, and the same dots carried by the transport map. The
cartogram is the picture in which the dots are uniform. Figure 2 draws
coastlines, borders and the 15-degree graticule through the transport
map and through the diffusion map of the same density.

\begin{figure}[htbp]\centering\includegraphics[width=\textwidth]{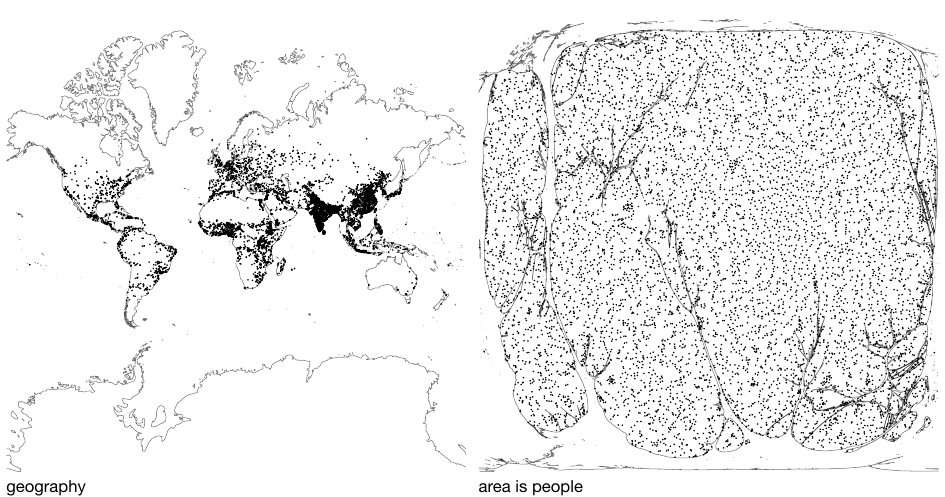}
\caption{One dot per million people (8,191 dots), placed by systematic sampling of the GHS-POP 2025 raster, on the Mercator frame (left) and under the optimal-transport map with land pure and a 20 per cent ocean margin (right). Coastlines are drawn through the same map.}\label{fig:dots}\end{figure}

\begin{figure}[htbp]\centering\includegraphics[width=\textwidth]{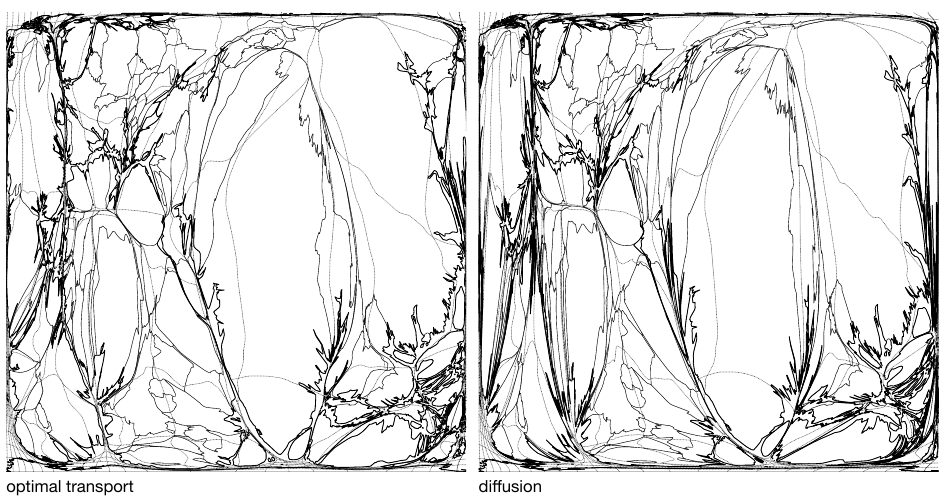}
\caption{Coastlines, national borders and the 15-degree graticule (dotted) through the transport map (left) and the diffusion map (right) of the same density: GHS-POP 2025 at 4096 cells, 30 km smoothing, a five per cent floor everywhere. Folds in empty cells were smoothed for drawing only.}\label{fig:lineart}\end{figure}

Table 1 gives the matched measurements at three grid sizes.

\begin{table}[htbp]\centering\small
\caption{Transport against diffusion on the same density (five per cent floor, smoothing 3 cells: 118 km at 1024, 60 km at 2048, 30 km at 4096). Density error: 5th and 95th population-weighted percentiles of $\rho/A - 1$. Anisotropy and rotation: population-weighted medians and 95th percentiles. Cost: population-weighted mean squared displacement in units of the frame width squared. Shape: population-weighted Procrustes residual over countries. Time: solve on an Apple M4.}\label{tab:matched}
\small\resizebox{\textwidth}{!}{\begin{tabular}{rlccccrr}
\toprule
grid & method & error p05 / p95 (\%) & anisotropy p50 / p95 & rotation p50 / p95 (deg) & cost ($10^{-3}$) & shape & time (s) \\
\midrule
1024 & OT & -2.5 / +2.1 & 3.65 / 18.9 & 0.0 / 0.0 & 81.96 & 0.753 & 6 \\
1024 & diffusion & -3.2 / +3.1 & 4.77 / 40.6 & 7.0 / 25.9 & 82.23 & 0.844 & 4 \\
2048 & OT & -2.4 / +2.2 & 3.81 / 21.9 & 0.0 / 0.0 & 82.05 & 0.773 & 26 \\
2048 & diffusion & -3.7 / +3.6 & 5.50 / 51.1 & 8.0 / 30.3 & 82.34 & 0.878 & 19 \\
4096 & OT & -2.5 / +2.4 & 3.93 / 25.1 & 0.0 / 0.0 & 81.98 & 0.785 & 131 \\
4096 & diffusion & -6.3 / +5.9 & 6.30 / 62.6 & 8.8 / 33.9 & 82.33 & 0.901 & 123 \\
\bottomrule
\end{tabular}}

\end{table}

Three things stand out. First, the transport cost is the same. At 4096
cells the diffusion map's population-weighted mean squared displacement
is within 0.4 per cent of the transport map's, and the
population-weighted mean displacement is 0.269 of the frame width for
transport and 0.269 for diffusion. The least-movement property of the
Brenier map, its defining virtue, is invisible in practice on this
problem: the diffusion flow is already nearly optimal in the mean-square
sense, because both maps carry the same bulk of population from the
continents into the same ocean. Second, both methods equalise the
smoothed density to within a few per cent: -2.5 to +2.4 per cent for
transport at every grid size, and -3.2 to +3.1 per cent for diffusion at
1024 cells widening to -6.3 to +5.9 per cent at 4096, where the advected
mesh accumulates more integration error. The transport solve is a direct
fixed point on the potential and its error is set by the residual
tolerance and the smoothing, not by the grid. Third, the maps differ in
how they distort, not in how far they move. The population-weighted
median local rotation is 8.8 degrees for diffusion and 34 degrees at the
95th percentile; for transport it is 0.01 and 0.05 degrees, zero up to
discretisation. The median anisotropy is 6.30 for diffusion against 3.93
for transport, and the 95th percentile 63 against 25. The Procrustes
shape error over countries is 0.901 against 0.785. Figure 3 shows the
two effects on Tissot indicatrices, and Figure 4 the full distributions.
The differences are consistent across the three grid sizes.

\begin{figure}[htbp]\centering\includegraphics[width=\textwidth]{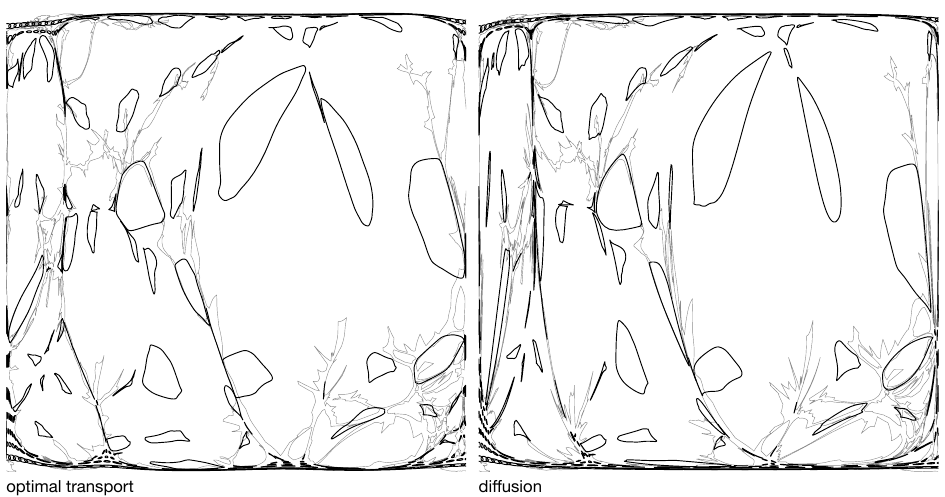}
\caption{Tissot indicatrices, geodesic circles of radius 300 km on a 15-degree lattice, through the transport map (left) and the diffusion map (right) of the same density. Shear shows as elongation, rotation as the turning of the ellipses' axes away from the graticule.}\label{fig:tissot}\end{figure}

\begin{figure}[htbp]\centering\includegraphics[width=\textwidth]{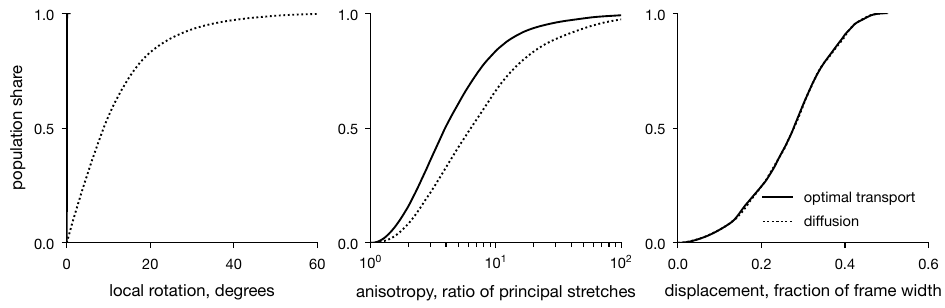}
\caption{Population-weighted cumulative distributions over source cells at 4096 cells: local rotation angle, anisotropy and displacement, transport (solid) against diffusion (dotted). The displacement curves coincide.}\label{fig:cdf}\end{figure}

The picture frame of Figures 1 and 2 was solved with land pure (\(s\) =
0.999) and the ocean at 20 per cent: density error -2.3 to +2.9 per
cent, median anisotropy 4.02, 301,535 folded cells out of 16,777,216,
all in the ocean creases, solved in 455 seconds.

The Delhi window (Figure 5), two degrees across with 69.8 million people
on the 3 arc-second raster, has a population-weighted density spread at
the 300 m scale of a factor 57 between its 5th and 95th percentiles
under the global map alone; after the local solve the spread is a factor
2.3, with the window's rim and everything outside it unchanged.

\begin{figure}[htbp]\centering\includegraphics[width=\textwidth]{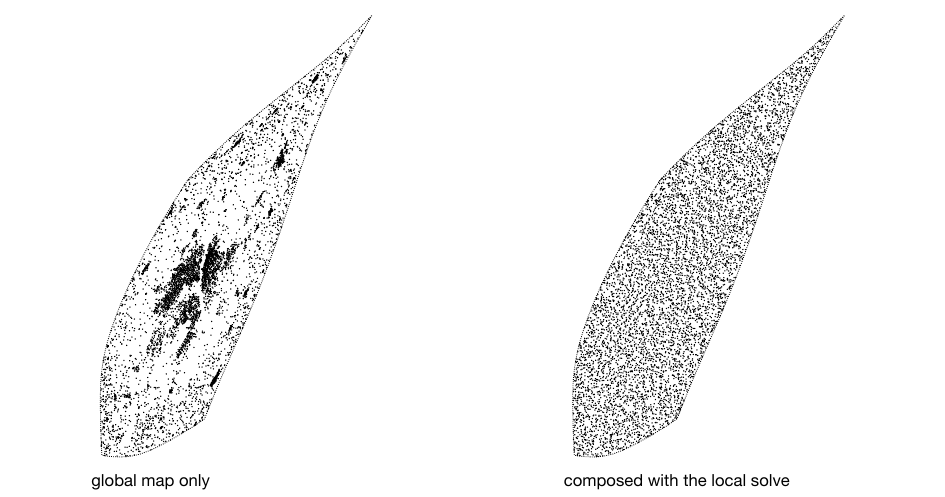}
\caption{The Delhi window, one dot per ten thousand people on the 3 arc-second raster, under the global 10 km map alone (left) and under the global map composed with the local 300 m transport solve (right). The dotted line is the window's rim, identical in both.}\label{fig:delhi}\end{figure}

The power diagram of humanity (Figure 6) partitions the same population
into 8,192 convex cells of 1.00 million people each; the mass of every
cell matches its target to a relative error below \(10^{-9}\).

\begin{figure}[htbp]\centering\includegraphics[width=\textwidth]{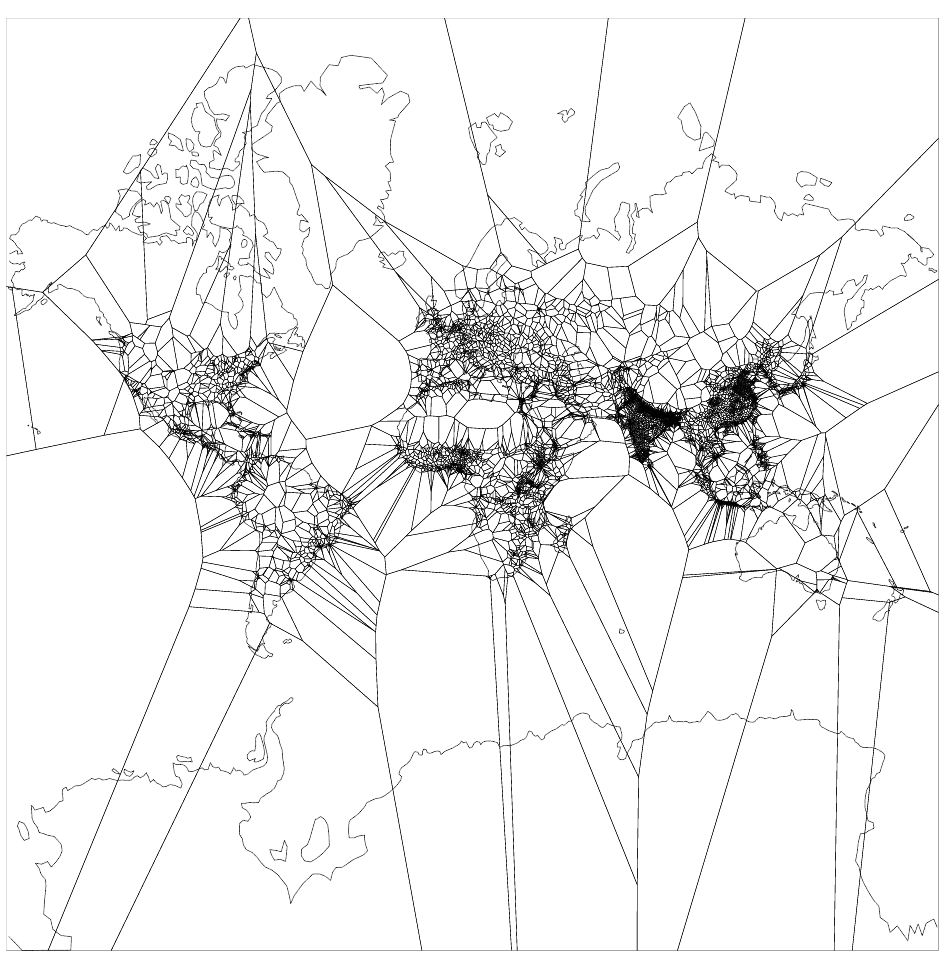}
\caption{Semi-discrete optimal transport from the population raster to 8,192 sites: every Laguerre cell holds 1.00 million people. The map is not deformed; the cells are. Coastline dotted.}\label{fig:power}\end{figure}

\FloatBarrier
\section{Discussion}\label{discussion}

What the transport map buys is not less movement but no rotation and
less shear. The population-weighted transport costs of the two methods
agree to well within one per cent, so a reader who wants the Brenier map
for its optimality gets nothing visible; a reader who wants coastlines
that keep their compass orientation and shapes that are stretched rather
than twisted gets a measurable difference, a median rotation of 9
degrees removed and a median anisotropy reduced by about 38 per cent.
Whether that difference is worth the fully nonlinear solve is a
judgement about the picture, and we have not measured readers.

Several limits bound the claims. The frame is a Mercator square, so the
construction is not sphere-native, unlike Weller and colleagues' meshes
or Sargent's sphere-optimised cartograms \citep{weller2016,sargent2024}; a periodic
equal-area variant of the same solver exists in the repository but is
not evaluated here. The 10 km smoothing sets the sharpness of the global
map and the 300 m smoothing that of the window; both are choices. The
floor and the ocean share are design parameters, as they are in every
method, and the pictures use a different pair from the measurements.
Folds remain at the ocean creases on the raw mesh at all sizes and are
repaired for drawing only. The comparison is against our own
implementation of the diffusion method rather than the authors' code,
and the flow-based method is absent. The solver reaches 4096 cells in
minutes on a laptop GPU, but Jacobs and Léger report the same grid size
in about a minute on a single CPU core with the back-and-forth method
\citep{jacobs2020}, so no speed claim is made.

Two observations may be useful beyond this construction. A diffusion
start time of half a pixel squared, which is common when a raster is
used unsmoothed, silently costs several per cent of density error and
should be replaced by pre-smoothing; and the local rotation of a
cartogram, which no standard quality measure reports \citep{nusrat2016,alam2015},
separates the methods more sharply than any area or shape statistic.

\section*{Data and code availability}\label{data-and-code-availability}
\addcontentsline{toc}{section}{Data and code availability}

The code, the experiment records with every parameter and metric, the
figure scripts with provenance files, and the manuscript source are
public in the repository \citep{repo}; the population raster is the
public GHS-POP R2023A release \citep{ghspop2023} and the boundaries are
Natural Earth \citep{naturalearth}.

\bibliographystyle{plainnat}
\bibliography{refs}

\end{document}